%% file: HIG-11-033_temp.tex
\pdfoutput=1

\documentclass[11pt,twoside,a4paper,cmspaper,final,collab]{cms-tdr}
\def\svnVersion{101375:102181}\def\cmsCernNo{2012-024}\def\cmsCernDate{\today}\def\cmsMessage{Submitted to Physics Letters B}

\begin{document}\cmsNoteHeader{HIG-11-033}

\hyphenation{had-ron-i-za-tion}
\hyphenation{cal-or-i-me-ter}
\hyphenation{de-vices}

\RCS$Revision: 101998 $
\RCS$HeadURL: svn+ssh://alverson@svn.cern.ch/reps/tdr2/papers/HIG-11-033/trunk/HIG-11-033.tex $
\RCS$Id: HIG-11-033.tex 101998 2012-02-07 11:08:40Z alverson $
\newlength\cmsFigWidth
\ifthenelse{\boolean{cms@external}}{\setlength\cmsFigWidth{0.45\textwidth}}{\setlength\cmsFigWidth{0.85\textwidth}}
\ifthenelse{\boolean{cms@external}}{\providecommand{\cmsLeft}{top}}{\providecommand{\cmsLeft}{left}}
\ifthenelse{\boolean{cms@external}}{\providecommand{\cmsRight}{bottom}}{\providecommand{\cmsRight}{right}}
\cmsNoteHeader{HIG-11-033} 
\title{Search for the standard model Higgs boson decaying into two photons in $\Pp\Pp$ collisions at $\sqrt{s}=7$ TeV}

\date{\today}

\abstract{A search for a Higgs boson decaying into two photons is described.
The analysis is performed using a dataset recorded by the CMS experiment at the
LHC from \Pp\Pp\ collisions at a centre-of-mass energy of 7\TeV, which corresponds
to an integrated luminosity of 4.8\fbinv.
Limits are set on the cross section of the standard model Higgs boson
decaying to two photons.
The expected exclusion limit at 95\% confidence level is between
1.4 and 2.4 times the standard model cross section in the mass range
between 110 and 150\GeV.
The analysis of the data excludes, at 95\% confidence level, the standard model
Higgs boson decaying into two photons in the mass range 128 to 132\GeV.
The largest excess of events above the expected standard model background is
observed for a Higgs boson mass hypothesis of 124\GeV with a local
significance of $3.1\,\sigma$.
The global significance of observing an excess with a local
significance ${\geq}3.1\,\sigma$ anywhere in
the search range 110--150\GeV is estimated to be $1.8\,\sigma$.
More data are required to ascertain the origin of this excess.
}
\hypersetup{%
pdfauthor={CMS Collaboration},%
pdftitle={Search for the standard model Higgs boson decaying into two photons in pp collisions at sqrt(s)=7 TeV},%
pdfsubject={Higgs search; two photon channel},%
pdfkeywords={CMS, physics, Higgs}}

\maketitle 

\newcommand{\ptgg}{\ensuremath{\PT^{\Pgg\Pgg}}\xspace}
\newcommand{\mgg}{\ensuremath{m_{\Pgg\Pgg}}\xspace}
\newcommand{\HqT}{{\textsc{h}q\textsc{t}}\xspace}
\newcommand{\mH}{\ensuremath{m_{\PH}\xspace}}
\section{Introduction}
\label{sec:intro}
The standard model
(SM)~\cite{SheldonL1961579,PhysRevLett.19.1264,SalamNobel}
of particle physics has been very successful in explaining experimental data.
The origin of the masses of the \PW\ and \cPZ\ bosons that arise from
electroweak symmetry breaking remains to be identified.
In the SM the Higgs mechanism is postulated, which leads to an additional
scalar field whose quantum, the Higgs boson, should be experimentally
observable~\cite{Englert:1964et,Higgs:1964ia,Higgs:1964pj,Guralnik:1964eu,Higgs:1966ev,Kibble:1967sv}.

Direct searches at the LEP experiments ruled out a SM Higgs boson lighter
than 114.4\GeV at 95\% confidence level (CL)~\cite{LEPHIGGS}.
Limits at 95\% CL on the SM Higgs boson mass have also been placed by
experiments at the Tevatron, excluding 162--166\GeV~\cite{TEVHIGGS_2010},
and by the ATLAS collaboration at the Large Hadron Collider (LHC),
excluding the ranges 145--206, 214--224,
and 340--450\GeV~\cite{ATLAS:2011aa, Aad:2011uq,ATLAS:2011af}.
Precision electroweak measurements indirectly constrain
the mass of the SM Higgs boson to be less
than 158\GeV at 95\% CL~\cite{EWWG:2010aa}.

The $\HGG$ decay channel provides a clean final-state topology with
a mass peak that can be reconstructed with high precision.
In the mass range 110~$< \mH <$ 150\GeV, $\HGG$ is one
of the more promising channels for a Higgs search at the LHC.
The primary production mechanism of the Higgs boson at the LHC is
gluon fusion with additional small contributions from vector boson
fusion (VBF) and production in association with a W or Z boson, or a
$\ttbar$ pair~\cite{Dawson:1990zj,Spira:1995rr,Harlander:2002wh,Anastasiou:2002yz,Ravindran:2003um,Actis:2008ug,Anastasiou:2008tj,Bolzoni:2010xr,deFlorian:2009hc,Ciccolini:2007jr,Ciccolini:2007ec,Han:1991ia}.
In the mass range 110~$< \mH <$ 150\GeV the SM $\HGG$ branching
fraction varies between 0.14\% and 0.23\%~\cite{Actis:2008ts}.
Previous searches in this channel have been conducted by the
CDF and D0 experiments~\cite{Aaltonen:2009ga,Abazov:2009kq}, and also
at the LHC by ATLAS~\cite{ATLAS:2011ww}.

This paper describes a search for a Higgs boson decaying into two photons in pp collisions
at a centre-of-mass energy of 7\TeV, using data taken in 2011 and
corresponding to an integrated luminosity of 4.8~\fbinv.
To improve the sensitivity of the search, selected diphoton events are
subdivided into classes according to indicators of mass resolution and
signal-to-background ratio.
Five mutually exclusive event classes are defined: four in terms of the pseudorapidity
and the shower shapes of the photons,
and a fifth class into which are put all events containing a pair of
jets passing selection requirements which are designed to select
Higgs bosons produced by the VBF process.

\section{The CMS detector}
\label{sec:detector}
A detailed description of the CMS detector can be found
elsewhere~\cite{bib-detector}.
The main features and those most pertinent to this analysis are described below.
The central feature is a superconducting solenoid,
13\unit{m} in length and 6\unit{m} in diameter,
which provides an axial magnetic field of 3.8\unit{T}.
The bore of the solenoid is instrumented with particle detection systems.
The steel return yoke outside the solenoid is instrumented with gas detectors
used to identify muons.
Charged particle trajectories are measured by the silicon pixel and
strip tracker,
with full azimuthal coverage within $|\eta| < 2.5$, where the pseudorapidity
$\eta$ is defined as $\eta = -\ln[\tan(\theta/2)]$, with $\theta$ being the
polar angle of the trajectory of the particle with respect to the
counterclockwise beam direction.
A lead-tungstate crystal electromagnetic calorimeter (ECAL) and a
brass/scintillator
hadron calorimeter (HCAL) surround the tracking volume and cover the region
$|\eta| < 3$.
The ECAL barrel extends to $|\eta| \approx$ 1.48.
A lead/silicon-strip preshower detector is located in front of the
ECAL endcap.
A steel/quartz-fibre Cherenkov forward calorimeter extends the
calorimetric coverage to $|\eta| <$ 5.0.
In the region $|\eta| < 1.74$, the HCAL cells have widths of 0.087
in both pseudorapidity and azimuth ($\phi$).
In the $(\eta, \phi)$ plane, and for $|\eta| < 1.48$, the HCAL cells map on
to $5\times5$
ECAL crystal arrays to form calorimeter towers projecting radially
outwards from points slightly offset from the nominal interaction point.
In the endcap, the ECAL arrays matching the HCAL cells contain fewer crystals.
Calibration of the ECAL uses $\Pgpz$s, $\PW\to\Pe\cPgn$, and $\cPZ\to
\Pe\Pe$.
Deterioration of transparency of the ECAL crystals due to irradiation
during the LHC
running periods and their subsequent recovery is monitored continuously
and corrected for using light injected from a laser and LED system.

\section{Data sample and reconstruction}
\label{sec:dataReco}

The dataset consists of events collected with diphoton triggers and
corresponds to an integrated luminosity of 4.8\fbinv.
Diphoton triggers with asymmetric transverse energy, \ET,
thresholds and complementary photon
selections were used.
One selection required a loose calorimetric
identification using the shower shape and very loose isolation requirements on photon
candidates, and the other required only that the
photon candidate had a high value of the $\RNINE$ variable.
This variable is defined as the energy sum of
$3\times 3$ crystals centred on the most energetic crystal in the
supercluster (described below) divided by the energy of the supercluster.
Its value is used in the analysis to identify photons undergoing a conversion.
The \ET thresholds used were at least
10\% lower than the final selection thresholds.
As the instantaneous luminosity delivered by the LHC increased,
it became necessary to tighten the isolation cut applied in the trigger.
To maintain high trigger efficiency, all four possible
combinations of threshold and selection criterion were deployed
(i.e., with both photon candidates having the $\RNINE$ condition,
with the high threshold candidate having the $\RNINE$ condition applied and
the low threshold candidate having the loose ID and isolation, and so on).
Accepting events that satisfy any of these triggers results in
a $>$99\% trigger efficiency for events passing the offline selection.

Photon candidates are reconstructed from clusters of ECAL channels around
significant energy deposits, which are merged into superclusters.
The clustering algorithms result in almost complete recovery of the energy
of photons that convert in the material in front of the ECAL.
In the barrel region, superclusters are formed from five-crystal-wide strips
in $\eta$ centred on the locally most energetic crystal (seed) and
have a variable extension in $\phi$.
In the endcaps, where the ECAL crystals do not have an $\eta \times \phi$
geometry, matrices of 5$\times$5 crystals (which may partially
overlap) around the most energetic crystals are merged if they lie within
a narrow road in $\eta$.

The photon energy is computed starting from the raw supercluster energy.
In the endcaps the preshower energy is added where the preshower is
present ($|\eta| >$ 1.65).
In order to obtain the best resolution, the raw energy is corrected
for the containment of the shower in the clustered crystals,
and the shower losses for photons which convert in
material upstream of the calorimeter.
These corrections are computed
using a multivariate regression technique based on the TMVA boosted
decision tree implementation~\cite{Hocker:2007ht}.
The regression is trained on photons in a sample of simulated events using the ratio of
the true photon energy to the raw energy as the target variable.
The input variables are the global
$\eta$ and $\phi$ coordinates of the supercluster, a collection of
shower-shape variables, and a set of local cluster coordinates.

Jets, used in the dijet tag, are reconstructed using a particle-flow
algorithm~\cite{CMS-PAS-PFT-09-001,CMS-PAS-PFT-10-002}, which
uses the information
from all CMS sub-detectors to reconstruct different types of particles produced in the event.
The basic objects of the particle-flow reconstruction
are the tracks of charged particles reconstructed in
the central tracker, and energy deposits reconstructed in the calorimetry.
These objects are clustered with the anti-k$_\mathrm{T}$ algorithm~\cite{Cacciari:2008gp}
using a distance parameter \DR = 0.5.
The jet energy measurement is calibrated to correct for detector effects
using samples of dijet, \GAMJET, and $\cPZ + \text{jet}$
events~\cite{Chatrchyan:2011ds}.
Energy from overlapping pp interactions other than that which produced the
diphoton (pile-up), and from the underlying event, is also included in
the reconstructed jets.
This energy is subtracted
using the \textsc{FastJet} technique~\cite{Cacciari:2007fd,Cacciari:2008gn,Cacciari:2011ma},
which is based on the calculation of the $\eta$-dependent transverse momentum density,
evaluated on an event-by-event basis.

Samples of Monte Carlo (MC) events used in the analysis
are fully simulated using \GEANTfour~\cite{Agostinelli:2002hh}.
The simulated events are reweighted to reproduce the distribution of
the number of interactions taking place in each bunch crossing.

\section{Vertex location}
\label{sec:vertex}

The mean number of \Pp\Pp\ interactions per bunch crossing
over the full dataset is 9.5.
The interaction vertices reconstructed using the tracks of charged particles
are distributed in the longitudinal direction, $z$, with an RMS spread of 6~cm.
If the interaction point is known to better than about 10\unit{mm}, then the
resolution on the opening angle between the photons makes a negligible
contribution to the mass resolution, as compared to the ECAL energy resolution.
Thus the mass resolution can be preserved by correctly assigning the
reconstructed photons to one of the interaction vertices reconstructed
from the tracks.
The techniques used to achieve this are described below.

The reconstructed primary vertex which most probably corresponds
to the interaction vertex of the diphoton event
can be identified using the kinematic properties
of the tracks associated with the vertex and their correlation with
the diphoton kinematics.
In addition, if either of the photons converts and the tracks from the conversion are
reconstructed and identified, the direction of the converted photon,
determined by combining the conversion vertex position and the position of the
ECAL supercluster, can be used to point to and so identify the diphoton
interaction vertex.

For the determination of the primary vertex position using kinematic
properties, three discriminating variables are
constructed from the measured scalar, $\pt$, or vector, $\vec{p}_\mathrm{T}$,
transverse momenta of the tracks associated with each vertex,
and the transverse momentum of the diphoton system,  $\ptgg$.
These three variables are: $\sum\pt^{2}$, and two
variables which quantify the $\pt$ balance with respect to the diphoton system:
-$\sum (\vec{p}_\mathrm{T}\cdot \frac{\vec{p}^{\Pgg\Pgg}_{T}}{|\vec{p}^{\Pgg\Pgg}_\mathrm{T}|})$
and  $(\sum\pt - \ptgg)/(\sum\pt + \ptgg)$.
An estimate of the pull to each vertex from the
longitudinal location on the beam axis pointed to by any reconstructed
tracks (from a photon conversion) associated with the two photon candidates is also
computed:
$|z_\text{conversion} - z_\text{vertex}|/\sigma_\text{conversion}$.
These variables are used in a multivariate system based on boosted
decision trees (BDT) to choose the reconstructed vertex to
associate with the photons.

The vertex-finding efficiency, defined as the efficiency to locate the vertex to within 10~mm of its
true position, has been studied with $\cPZ\to\Pgm\Pgm$
events where the algorithm is run after the removal of the muon
tracks.
The use of tracks from a converted photon to locate the vertex is studied with
$\Pgg+\text{jet}$ events.
In both cases the ratio of the efficiency measured in data to that
in MC simulation is close to unity.
The value is measured as a function of the boson $\pt$, as measured by
the reconstructed muons, and is used as a correction to the Higgs
boson signal model.
An uncertainty of 0.4\% is ascribed to the knowledge of the vertex finding
efficiency coming from
the statistical uncertainty in the efficiency measurement from
$\cPZ\to\Pgm\Pgm$ (0.2\%) and the uncertainty related to the Higgs boson \pt
spectrum description, which is estimated to be 0.3\%.
The overall vertex-finding efficiency for a
Higgs boson of mass 120\GeV, integrated over its \pt spectrum, is
computed to be 83.0$\pm$0.2(stat)$\pm$0.4(syst)\%.

\section{Photon selection}
\label{sec:selection}

The event selection requires two photon candidates with
$\pt^{\Pgg}(1) > \mgg/3$ and $\pt^{\Pgg}(2) > \mgg/4$
within the ECAL fiducial region, $|\eta|<2.5$, and excluding the
barrel-endcap transition region $1.44 < |\eta| < 1.57$.
The fiducial region requirement  is applied to the supercluster position in the ECAL,
and the $\pt$ threshold is applied after the vertex assignment.
The excluded barrel-endcap transition region removes from the
acceptance the last two
rings of crystals in the barrel, to ensure complete containment
of accepted showers, and the first ring of trigger towers in the endcap which
is obscured by cables and services exiting between the barrel and endcap.
In the rare case where the event contains more than two photons
passing all the selection requirements, the pair with the highest summed
(scalar) \pt is chosen.

The dominant backgrounds to $\HGG$ consist of 1) the irreducible background
from the prompt diphoton  production, and 2) the reducible
backgrounds from $\Pp\Pp\to\GAMJET$ and
$\Pp\Pp\to\text{jet}+\text{jet}$ where one or more of the
``photons'' is not a prompt photon.
Photon identification requirements are used to greatly reduce the contributions
from non-prompt photon background.

Isolation is a powerful tool to reject the non-prompt
background due to electromagnetic showers originating in jets -- mainly
due to single and multiple $\Pgpz$s.
The isolation of the photon candidates is measured by summing the
transverse momentum (or energy) found in the tracker, ECAL or HCAL
within a distance $\Delta$R = $\sqrt{\Delta\eta^{2} + \Delta\phi^{2}}$
of the candidate (values of $\Delta$R = 0.3 and $\Delta$R = 0.4 are
used).
The tracks or calorimeter energy deposits very close to the candidates,
which might originate from the candidate itself, are excluded from the sum.
Pile-up results in two complications.
First, the $\et$ summed in the isolation region in the ECAL and in the
HCAL includes a contribution from other collisions
in the same bunch crossing.
The isolation sums in the ECAL and HCAL, and hence both the efficiency
and rejection power of selection based on the sums, are thus dependent on the
number of interactions in the bunch crossing.
Second, the track isolation
requires that the tracks used in the isolation sum are matched to the
chosen vertex (so that the sum does not suffer from pile-up).
If the vertex is incorrectly assigned, the isolation sum will
be unrelated to the true isolation of the candidate.
This allows non-prompt candidates which are not, in fact,
isolated from tracks originating from their interaction point, to appear
isolated.

The first issue is dealt with by calculating the median
transverse energy density in the event, $\rho$, in regions of the detector
separated from the jets and photons, and subtracting an appropriate amount,
proportional to $\rho$, from the isolation sums.
The second problem is dealt with by applying a selection requirement not only on the isolation
sum calculated using the chosen diphoton vertex, but also on the isolation
sum calculated using the vertex hypothesis which maximises the sum.
The isolation requirements are applied as a constant fraction of the candidate photon
$\pt$, effectively cutting harder on low $\pt$ photons.
It has been shown with $\cPZ\to\Pe\Pe$ events that the resulting variation of
selection efficiency with $\pt$ is well modelled in the simulation.

In addition to isolation variables, the following observables are also
used for photon selection: the ratio of hadronic energy behind the photon to the photon
energy, the transverse width of the
electromagnetic shower, and an electron track veto.

Photon candidates with high values of $\RNINE$ are mostly unconverted
and have less background than those with lower values.
Photon candidates in the barrel have less background than those in the
endcap.
For this reason it has been found useful to divide photon candidates into
four categories and apply a different selection in each category,
using more stringent requirements in categories with higher background
and worse resolution.

\begin{table*}[htbp]
\caption{Photon identification efficiencies measured in the four photon
categories using a tag and probe technique applied to $\cPZ\to\Pe\Pe$
events (for all requirements except the electron veto).
Both statistical and systematic errors are given for the data
measurement (in that order), and these are combined quadratically to
calculate error on the ratio $\epsilon_{\text{data}}$/$\epsilon_{\text{MC}}$.}
\begin{center}
\begin{tabular}{|l|c|c|c|}
\hline
Category &
$\epsilon_\text{data}$ (\%) & $\epsilon_{\text{MC}}$ (\%) & $\epsilon_{\text{data}}$/$\epsilon_{\text{MC}}$ \\
\hline
Barrel, $\RNINE >$ 0.94 & 89.26$\pm$0.06$\pm$0.04 & 90.61$\pm$0.05 & 0.985$\pm$0.001 \\
Barrel, $\RNINE <$ 0.94 & 68.31$\pm$0.06$\pm$0.55 & 68.16$\pm$0.05 & 1.002$\pm$0.008 \\
Endcap, $\RNINE >$ 0.94 & 73.65$\pm$0.14$\pm$0.39 & 73.45$\pm$0.12 & 1.002$\pm$0.006 \\
Endcap, $\RNINE <$ 0.94 & 51.25$\pm$0.11$\pm$1.25 & 48.70$\pm$0.08 & 1.052$\pm$0.026 \\
\hline
\end{tabular}
\end{center}
\label{tab:PhotEff}
\end{table*}

The efficiency of the photon identification is measured in data using
tag and probe techniques~\cite{springerlink:10.1007/JHEP10(2011)132}.
The efficiency of the complete selection excluding the electron veto
requirement is determined using $\cPZ\to\Pe\Pe$ events.
Table~\ref{tab:PhotEff} shows the results for data and MC
simulation, and the ratio of efficiency in data to that in the
simulation, $\epsilon_\text{data}$/$\epsilon_\text{MC}$.
The efficiency for photons to pass the electron veto has been measured
using $\cPZ\to\Pgm\Pgm\Pgg$ events, where the photon is produced by final-state
radiation, which provide a rather pure source of prompt photons.
The efficiency approaches
100\% in all except the fourth category, where it is 92.6$\pm$0.7\%,
due to imperfect pixel detector coverage at large $\eta$.
The ratio $\epsilon_\text{data}$/$\epsilon_\text{MC}$ for the electron veto
is close to unity in all categories.
The quadratic sum of the statistical and systematic uncertainties
for the measurements of efficiencies using data are propagated to the
uncertainties on the ratios.
The ratios are used as corrections to the signal efficiency simulated in the
MC model of the signal.
The uncertainties on the ratios are taken as a systematic
uncertainties in the limit setting.

The efficiency of the trigger has also been measured using $\cPZ\to\Pe\Pe$
events, with the events classified as described below.
For events passing the analysis selection the trigger efficiency is
found to be 100\% in the high $\RNINE$ event classes, and about 99\%
in the other two classes.

\section{Event classes}
\label{sec:classes}

The sensitivity of the search can be enhanced by subdividing the
selected events into classes
according to indicators of mass resolution and signal-to-background ratio
and combining the results of a search in each class.

Two photon classifiers are used: the minimum $\RNINE$ of the two photons,
$\RNINE^\text{min}$, and the maximum pseudorapidity (absolute value) of the two photons,
giving four classes based on photon properties.
The class boundary values for $\RNINE$ and pseudorapidity are the
same as those used to categorize photon candidates for the photon
identification cuts.
These photon classifiers are effective in separating diphotons whose
mass is reconstructed with good resolution from those whose mass is
less well measured and in separating events for which the
signal-to-background probability is higher from those for which it is lower.

A further class of events includes any event passing
a dijet tag defined to select Higgs bosons produced by
the VBF process.
Events in which a Higgs boson is produced by VBF have two forward jets,
originating from the two scattered quarks.
Higgs bosons produced by this mechanism have a harder
transverse momentum spectrum than those
produced by the gluon-gluon fusion process or the photon pairs
produced by the background processes~\cite{Ballestrero:2008gf}.
By using a dijet tag it is possible to define a small class of events
which have an expected signal-to-background ratio more than an order of
magnitude greater than events in the four classes defined by photon properties.
The additional classification of events into a
dijet-tagged class improves the sensitivity of the analysis by about 10\%.

Candidate diphoton events for the dijet-tagged class have the same
selection requirements imposed on the photons as for the other classes with the exception of
the $\pt$ thresholds, which are modified to increase signal acceptance.
The threshold requirements for this class are
$\pt^{\Pgg}(1)~> 55\times\mgg/120$, and $\pt^{\Pgg}(2)~> 25\GeV$.

The selection variables for the jets use the two highest transverse
energy ($\ET$) jets in the event with pseudorapidity $|\eta|<$ 4.7.
The pseudorapidity restriction with respect to the full calorimeter
acceptance ($|\eta| < $5), avoids the use of jets for which
the energy corrections are less reliable and
is found to have only a small effect ($< $2\% change) on the signal efficiency.
The following selection requirements have been optimized
using simulated events, of VBF signal and diphoton background, to
improve the expected limit at 95\% CL on the VBF signal cross section,
using this class of events alone.
The $\ET$ thresholds for the two jets are 30 and 20\GeV, and the pseudorapidity
separation between them is required to be greater than 3.5.
Their invariant mass is required to be greater than 350\GeV.
Two additional selection criteria, relating the dijet to the diphoton
system, have been applied: the difference
between the average pseudorapidity of the two jets and the pseudorapidity
of the diphoton system is required to be less than 2.5~\cite{Rainwater:1996ud},
and the difference in azimuthal angle between the diphoton system
and the dijet system is required to be greater than 2.6 radians
($\approx$150\de).

For a Higgs boson having a mass, \mH, of 120\GeV the overall
acceptance times selection efficiency of the dijet tag for
Higgs boson events is 15\% (0.5\%) for those produced by VBF (gluon-gluon fusion).
This corresponds to about 2.01 (0.76) expected events.
Events passing this tag are excluded from the four classes defined
by $\RNINE$ and pseudorapidity, but enter the fifth class.
About 3\% of Higgs boson signal events are expected to be removed from
the four classes defined by diphoton properties.
In the mass range 100 $<\mgg<$ 180\GeV the fractions of diphoton events
in the selected data, which pass the dijet VBF tag and enter the fifth class,
and which would otherwise have entered one of the four classes defined
in Table~\ref{tab:ClassFracs}, are 0.8\%, 0.5\%, 0.3\% and 0.4\%,
respectively.

\begin{table*}[htbp]
\begin{center}
\caption{Number of selected events in different event classes,
for a SM Higgs boson signal ($\mH=120\GeV$), and for data at 120\GeV.
The value given for data, expressed as events/GeV, is obtained by dividing
the number of events in a bin of $\pm$ 10\GeV, centred at 120\GeV, by 20\GeV.
The mass resolution for a SM Higgs boson signal in each
event class, is also given.}
\begin{tabular}{|l|c|c|c|c|c|}
\hline
\multirow{2}{*}{} & \multicolumn{2}{c|}{Both photons in barrel} & \multicolumn{2}{c|}{One or both in endcap} & Dijet \\
\cline{2-5}
 & $\RNINE^\text{min}>$0.94 & $\RNINE^\text{min}<$0.94 & $\RNINE^\text{min}>$0.94 & $\RNINE^\text{min}<$0.94 & tag \\
\hline
SM signal expected & 25.2 (33.5\%) & 26.6 (35.3\%) & 9.5 (12.6\%) & 11.4 (14.9\%) & 2.8 (3.7\%) \\
Data (events/\GeVns) & 97.5 (22.8\%) & 143.4 (33.6\%) & 76.7 (17.9\%) &
107.4 (25.1\%) & 2.3 (0.5\%) \\
\hline
$\sigma_\text{eff}$ (\GeVns) & 1.39 & 1.84 & 2.76 & 3.19 & 1.71 \\
FWHM/2.35 (\GeVns)                & 1.19 & 1.53 &  2.81 & 3.18 & 1.37\\
\hline
\end{tabular}
\label{tab:ClassFracs}
\end{center}
\end{table*}

The number of events in each of the five classes is shown in
Table~\ref{tab:ClassFracs}, for signal events from all Higgs
boson production processes (as predicted by MC simulation), and for data.
A Higgs boson with \mH=120\GeV is chosen for the signal, and the
data are counted in a bin ($\pm$ 10\GeV) centred at 120\GeV.
The table also shows the mass resolution, parameterized both as
$\sigma_\text{eff}$,
half-the-width of the narrowest window containing 68.3\% of the
distribution, and as the full width at half maximum (FWHM) of the invariant
mass distribution divided by 2.35.
The resolution in the endcaps is noticeably worse than in the barrel
due to several factors, which include the amount of material
in front of the calorimeter and less precise single channel
calibration.

Significant systematic uncertainties on the efficiency of dijet tagging of
signal events arise from the uncertainty on the MC modelling of jet-energy
corrections and jet-energy resolution, and from uncertainties in predicting
the presence of the jets and their kinematics.
These uncertainties arise from the effect of different underlying event tunes,
and from the uncertainty on parton distribution functions and QCD scale factor.
Overall, an uncertainty of 10\% is assigned to the efficiency
for VBF signal events to enter the dijet tag class, and an
uncertainty of 70\%, which is dominated by the uncertainty on the
underlying event tune is assigned to the efficiency for signal events
produced by gluon-gluon fusion to enter the dijet-tag class.
The uncertainty on the underlying event tunes was investigated
by comparing the \textsc{dt}6~\cite{Bartalini:2010su},
\textsc{p}0~\cite{Skands:2010ak},
\textsc{p}ro\textsc{pt}0 and \textsc{p}ro\textsc{q}20~\cite{Buckley:2009bj} tunes
to the \textsc{z}2 tune~\cite{Field:2010bc} in \PYTHIA~\cite{Sjostrand:2006za}.

\section{Background  and signal modelling}
\label{sec:modelling}

The MC simulation of the background processes is not used in the analysis.
However, the diphoton mass spectrum that is observed after the full
event selection is found to agree with the distribution predicted by
MC simulation, within the uncertainties on
the cross sections of the contributing processes which is estimated to
be about 15\%.
The background components have been scaled by
$K$-factors obtained from CMS measurements~\cite{CMS-DY,CMS-gg,CMS-jj}.
The contribution to the background in the
diphoton mass range $110< \mgg <150\GeV$ from
processes giving non-prompt photons is about 30\%.

The background model is obtained by fitting
the observed diphoton mass distributions in each of the five event classes
over the range $100< \mgg <180\GeV$.
The choice of function used to fit the background, and the choice of
the range, was made based on a study of the possible bias introduced by the choice
on both the limit, in the case of no signal,
and the measured signal strength, in the case of a signal.

The bias studies were performed using background-only and
signal-plus-background MC simulation samples and showed that for the first four
classes, the bias in either excluding or finding a
Higgs boson signal in the mass range $110< \mgg <150\GeV$ can be ignored,
if a 5$^\mathrm{th}$ order polynomial fit
to the range 100~$< \mgg <$~180\GeV is used.
In both cases the maximum bias was found to be at
least five times smaller than the statistical uncertainties of the fit.
For the dijet-tagged event class, which contains much fewer events, the use of
a 2$^\mathrm{nd}$ order polynomial was shown to be sufficient and unbiased.

The description of the Higgs boson signal used in the
search is obtained from MC simulation using the
next-to-leading order (NLO) matrix-element generator
\POWHEG~\cite{powheg1,powheg2} interfaced with
\PYTHIA~\cite{Sjostrand:2006za}, using the \textsc{z}2 underlying event tune.
For the dominant gluon-gluon fusion process, the Higgs boson
transverse momentum spectrum has been reweighted to the
next-to-next-to-leading logarithmic (NNLL) + NLO distribution computed
by the \HqT program~\cite{HqT1,HqT2,deFlorian:2011xf}.
The uncertainty on the signal cross section
due to PDF uncertainties has been determined using the PDF4LHC
prescription~\cite{Botje:2011sn,Alekhin:2011sk,Lai:2010vv,Martin:2009iq,Ball:2011mu}.
The uncertainty on the cross section due to scale uncertainty has
been estimated by varying  independently the renormalization and factorization scales
used by \HqT, between \mH/2 and 2\mH.
We have verified that the effect of this variation on the rapidity of
the Higgs boson is very small and can be neglected.

Corrections are made to the measured energy of the photons based
on detailed study of the mass distribution of $\cPZ\to\Pe\Pe$ events and
comparison with MC simulation.
After the application of these corrections the $\cPZ\to\Pe\Pe$ events
are re-examined and values are derived for the
random smearing that needs to be made to the MC simulation
to account for the energy resolution observed in the data.
These smearings are derived for photons separated into four $\eta$ regions
(two in the barrel and two in the endcap) and two categories of $\RNINE$.
The uncertainties on the measurements of the photon scale and
resolution are taken as systematic uncertainties
in the limit setting.
The overall uncertainty on the diphoton mass scale is less than 1\%.

\begin{figure*}[htbp]
   \begin{center}
      \includegraphics[width=0.99\linewidth]{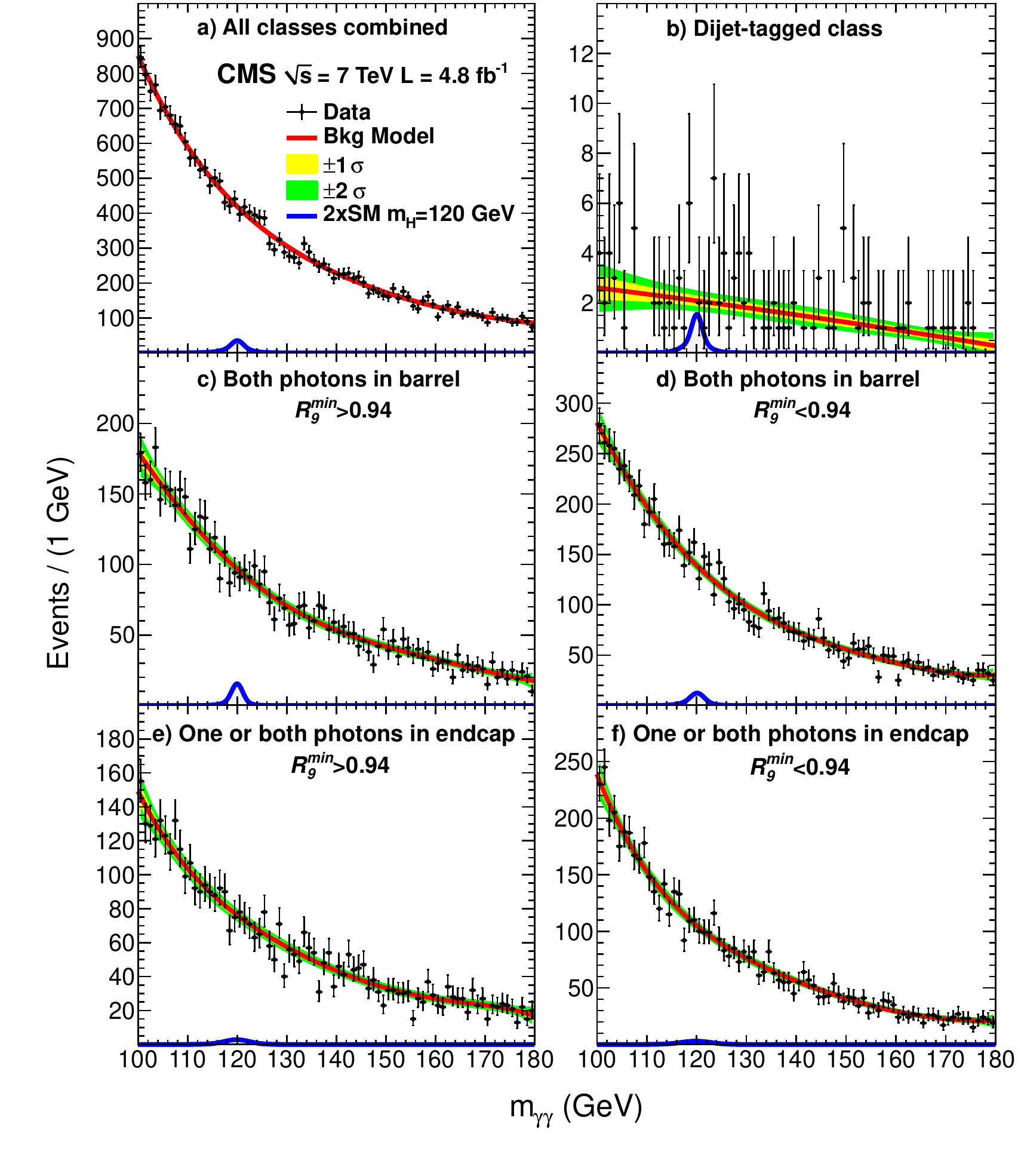}
     \caption{\label{fig:BckSig}Background model fit to the \mgg
distribution for the five event classes, together with a simulated signal
($\mH$=120\GeV). The magnitude of the simulated signal is what would
be expected if its cross section were twice the SM expectation.
The sum of the event classes together with the sum of the five fits is
also shown.
a) The sum of the five event classes.
b) the dijet-tagged class,
c) both photons in the barrel, $\RNINE^\text{min} >$ 0.94,
d) both photons in the barrel, $\RNINE^\text{min} <$ 0.94,
e) at  least one photon in the endcaps, $\RNINE^\text{min} >$ 0.94,
f) at  least one photon in the endcaps, $\RNINE^\text{min} <$ 0.94.
}
   \end{center}
\end{figure*}

The \mgg distributions for the data in the five event classes,
together with the background fits, are shown in Fig.~\ref{fig:BckSig}.
The uncertainty bands shown are computed from the fit
uncertainty on the
background yield within each bin used for the data points.
The expected signal shapes for $\mH=120\GeV$ are also shown.
The magnitude of the simulated signal is what would be expected if its cross section
were twice the SM expectation.
The sum of the five event classes is also shown, where the line
representing the background model is the sum of the five fits to the
individual event classes.

\section{Results}
\label{sec:results}
The confidence level for exclusion or discovery of a SM Higgs boson
signal is evaluated
using the diphoton invariant mass distribution for each of
the event classes.
The results in the five classes are combined in the CL calculation to obtain the final result.

The limits are evaluated using
a modified frequentist approach, CL$_\mathrm{s}$,
taking the profile likelihood as a test statistic~\cite{Junk:1999kv,Read:2000ru,ATLAS:1379837}.
Both a binned and an unbinned evaluation of the likelihood are considered.
While most of the analysis and determination of systematic uncertainties
are common for these two approaches,
there are differences at the final stages which make a comparison
useful.
The signal model is taken from MC simulation after applying the
corrections determined from data/simulation comparisons of $\cPZ\to\Pe\Pe$ and
$\cPZ\to\Pgm\Pgm\Pgg$ events mentioned above,
and the reweighting of the Higgs boson transverse momentum spectrum.
The background is evaluated from a fit to the data without reference
to the MC simulation.

Since a Higgs boson signal would be reconstructed with a mass resolution approaching
1\GeV in the classes with best resolution, the limit and signal
significance evaluation is carried out in steps of 0.5\GeV.
The SM Higgs boson cross sections and branchings ratios used are taken
from ref.~\cite{LHCHiggsCrossSectionWorkingGroup:2011ti}.

Table~\ref{tab:systematics} lists the sources of systematic uncertainty
considered in the analysis, together with
the magnitude of the variation of the source that has been applied.

\begin{table*}[htbp]
\begin{center}
\caption{Separate sources of systematic uncertainties accounted for in
this analysis. The magnitude of the variation of the source that has
been applied to the signal model is shown in the second column.}
\begin{tabular}{| l r|c|c|}
\hline
\multicolumn{2}{| l |}{\textbf{Sources of systematic uncertainty}} & \multicolumn{2}{ c |}{\textbf{Uncertainty}}\\
\hline
\hline
\multicolumn{2}{| l |}{\textbf{Per photon}} & Barrel & Endcap \\
\hline
\multicolumn{2}{| l |}{Photon identification efficiency} & 1.0\% & 2.6\%\\
\multicolumn{2}{| l |}{$\RNINE >$0.94 classification (class migration)} & 4.0\% & 6.5\%\\
Energy resolution ($\Delta\sigma/E_\text{MC}$) & $\RNINE > 0.94$ (low $\eta$, high $\eta$) & 0.22\%, 0.61\% & 0.91\%, 0.34\% \\
& $\RNINE < 0.94$ (low $\eta$, high $\eta$) & 0.24\%, 0.59\% & 0.30\%, 0.53\% \\
Energy scale ($(E_\text{data}-E_\text{MC})/E_\text{MC}$) & $\RNINE > 0.94$ (low $\eta$, high $\eta$) & 0.19\%, 0.71\% & 0.88\%, 0.19\% \\
& $\RNINE < 0.94$ (low $\eta$, high $\eta$) & 0.13\%, 0.51\% & 0.18\%, 0.28\% \\
\hline
\hline
\multicolumn{4}{| l |}{\textbf{Per event}}\\
\hline
\multicolumn{2}{|l|}{Integrated luminosity} & \multicolumn{2}{ c |}{4.5\%} \\
\multicolumn{2}{|l|}{Vertex finding efficiency} & \multicolumn{2}{ c |}{0.4\%}\\
\multicolumn{2}{|l|}{Trigger efficiency \hfill One or both photons $\RNINE <$ 0.94 in endcap} & \multicolumn{2}{ c |}{0.4\%} \\
\multicolumn{2}{|r|}{Other events} & \multicolumn{2}{ c |}{0.1\%} \\
\hline
\hline
\multicolumn{4}{| l |}{\textbf{Dijet selection}}\\
\hline
Dijet-tagging efficiency & VBF process & \multicolumn{2}{ c |}{10\%}\\
 & Gluon-gluon fusion process & \multicolumn{2}{ c |}{70\%}\\
\hline
\hline
\multicolumn{2}{| l |}{\textbf{Production cross sections}} & Scale & PDF \\
\hline
\multicolumn{2}{|l|}{Gluon-gluon fusion} & +12.5\% -8.2\% & +7.9\% -7.7\% \\
\multicolumn{2}{|l|}{Vector boson fusion} & +0.5\% -0.3\% & +2.7\% -2.1\% \\
\multicolumn{2}{|l|}{Associated production with $\PW/\cPZ$} & 1.8\% & 4.2\% \\
\multicolumn{2}{|l|}{Associated production with $\ttbar$} & +3.6\% -9.5\% & 8.5\% \\
\hline
\end{tabular}
\label{tab:systematics}
\end{center}
\end{table*}

\begin{figure}[htbp]
   \begin{center}
      \includegraphics[width=\cmsFigWidth]{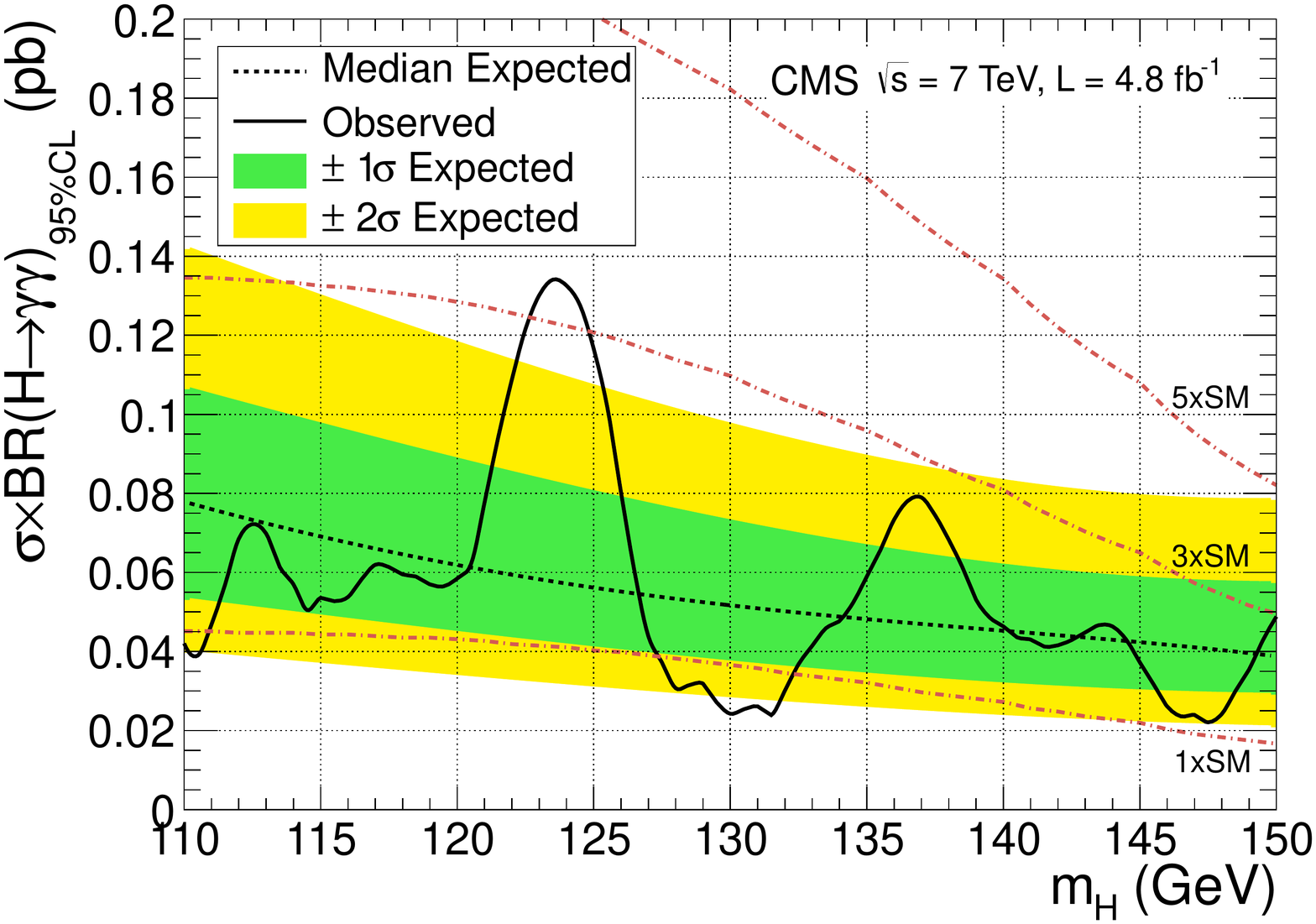}
      \includegraphics[width=\cmsFigWidth]{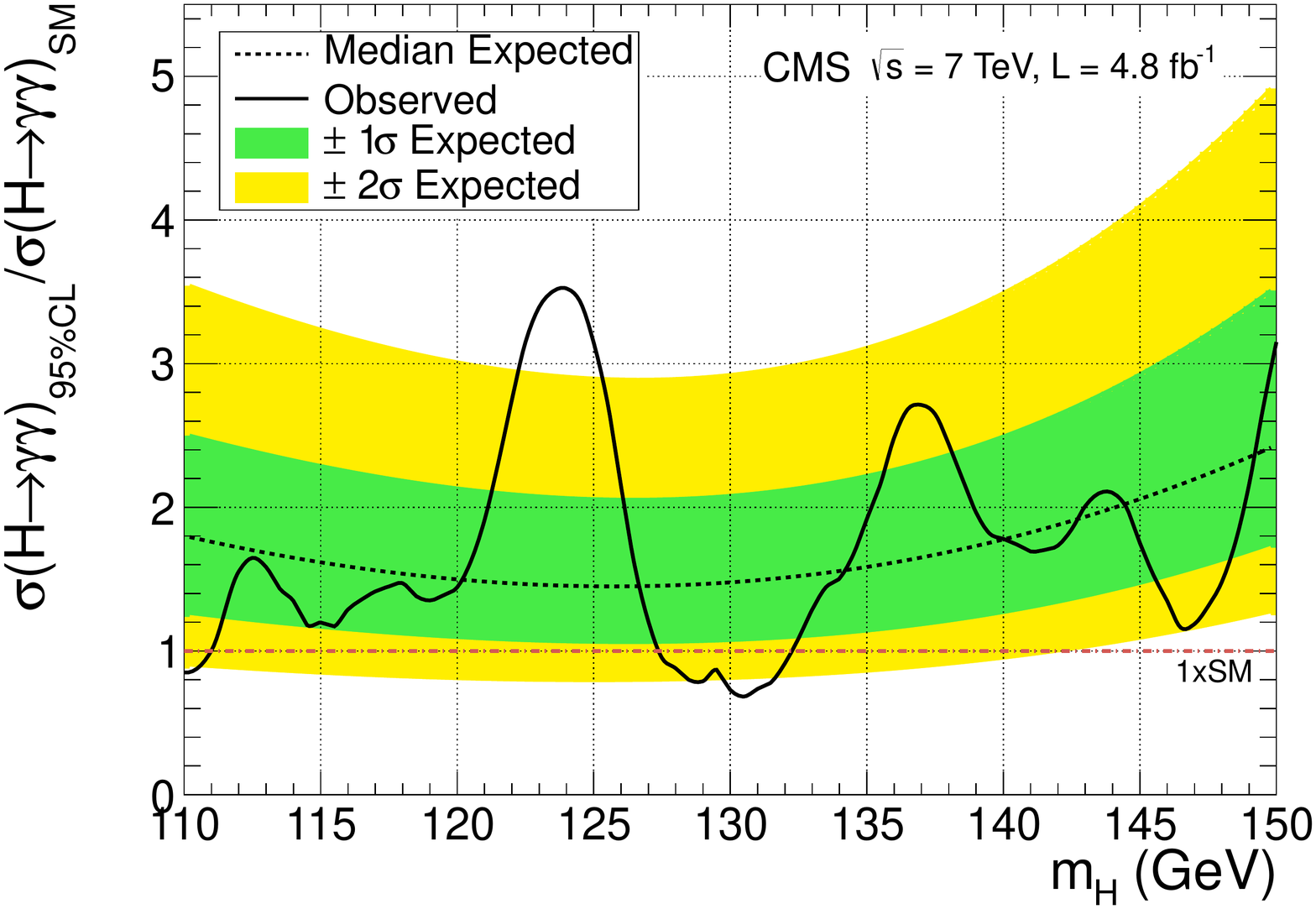}
      \caption{\label{fig:SMlimit}Exclusion limit on the cross section
        of a SM Higgs boson decaying into two photons as a function of
        the boson mass (upper plot). Below is the same exclusion limit
        relative to the
        SM Higgs boson cross section, where the theoretical uncertainties on the
        cross section have been included in the limit setting.}
   \end{center}
\end{figure}

The limit set on the cross section of a Higgs boson decaying to two photons
using the frequentist CL$_\mathrm{S}$ computation
and an unbinned evaluation of the likelihood,
is shown in Fig.~\ref{fig:SMlimit}.
Also shown is the limit relative to the SM expectation,
where the theoretical uncertainties on the expected cross sections
from the different production mechanisms are individually included
as systematic uncertainties in the limit setting procedure.
The observed limit excludes at 95\% CL the standard model
Higgs boson decaying into two photons in the mass range 128 to 132\GeV.
The fluctuations of the observed limit about the expected limit are
consistent with statistical fluctuations to be expected in
scanning the mass range.
The largest deviation, at \mgg=124\GeV, is discussed in more detail below.
It has also been verified that the shape of the observed limit
is insensitive to the choice of background model fitting function.
The results obtained from the binned evaluation of the likelihood are
in excellent agreement with the results shown in Fig.~\ref{fig:SMlimit}.

\begin{figure}[htbp]
   \begin{center}
      \includegraphics[width=\cmsFigWidth]{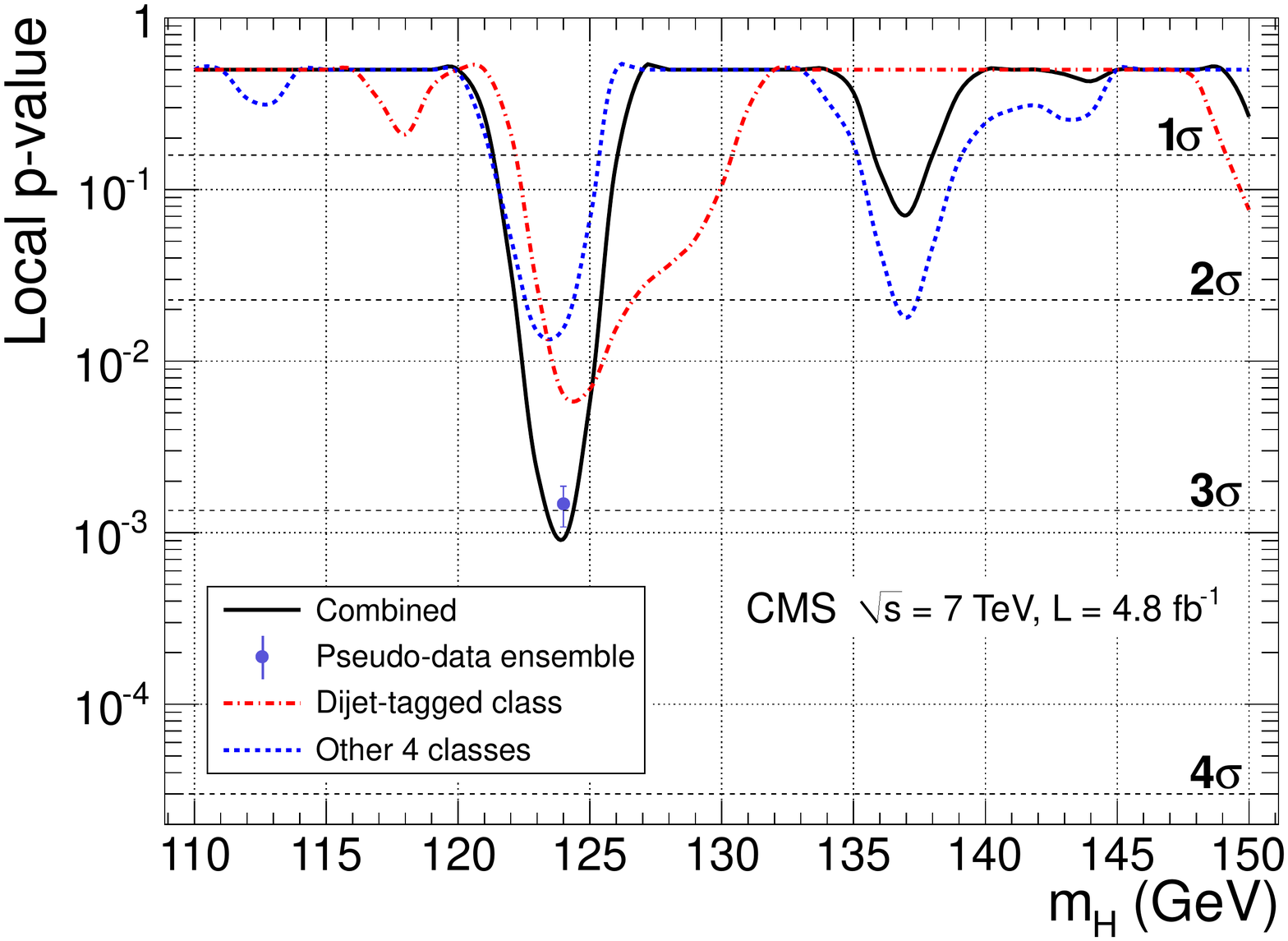}
      \caption{\label{fig:pvalue}The local $p$-value as a function of
        Higgs boson mass, calculated in the asymptotic approximation.
The point at 124\GeV shows the value obtained with a pseudo-data ensemble.}
   \end{center}
\end{figure}

\begin{figure*}[htbp]
   \begin{center}
      \includegraphics[width=0.85\textwidth]{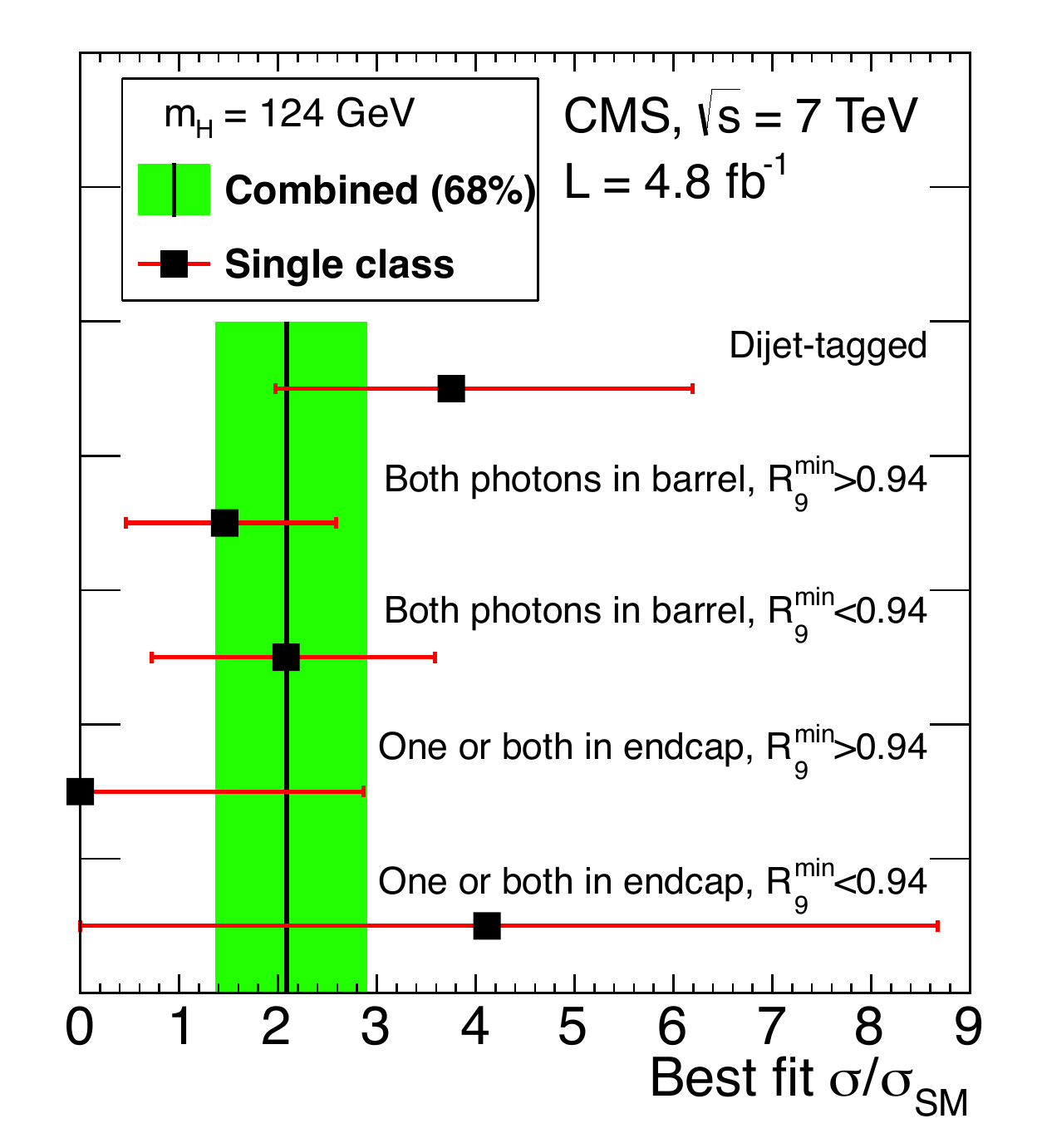}
      \caption{\label{fig:mu}The best fit signal strength, in terms of
        the standard model Higgs boson cross section, for the combined
        fit to the five classes (vertical line) and for the individual
        contributing classes (points) for the hypothesis of a SM Higgs
        boson mass of 124\GeV. The band corresponds to $\pm 1\,\sigma$
uncertainties on the overall value. The horizontal bars indicate
$\pm 1\,\sigma$ uncertainties on the
values for individual classes.
}
   \end{center}
\end{figure*}

Figure~\ref{fig:pvalue} shows the local $p$-value calculated,
using the asymptotic approximation~\cite{Cowan:2010st}, at 0.5\GeV
intervals in the mass range $110<\mH<150\GeV$.
The local $p$-values for the dijet-tag event class, and for the
combination of the four other classes, are also
shown (dash-dotted and dashed lines respectively).
The local $p$-value quantifies the probability for the background to
produce a fluctuation at least as large as observed, and
assumes that the relative signal strength between the
event classes follows the MC signal model for the standard model Higgs boson.
The local $p$-value corresponding to the largest upwards fluctuation
of the observed limit, at 124\GeV, has been computed to be
9.2$\times$10$^{-4}$ ($3.1\,\sigma$) in the asymptotic approximation, and
1.5$\pm$0.4$\times10^{-3}$ ($3.0\,\sigma$) when the calculation uses pseudo-data
(the value for the pseudo-data ensemble at 124\GeV is shown in Fig.~\ref{fig:pvalue}).
The combined best fit signal strength, for a SM Higgs boson mass hypothesis
of 124\GeV, is 2.1$\pm$0.6 times the SM Higgs boson cross section.
In Fig.~\ref{fig:mu} this combined best fit signal strength is compared
to the best fit signal strengths in each of the event classes.
Since a fluctuation of the background could occur at any point in the mass
range there is a look-elsewhere effect~\cite{LEE}.
When this is taken into account the probability, under the background
only hypothesis, of observing a similar or larger excess in the full
analysis mass range ($110<\mH<150\GeV$)
is 3.9$\times$10$^{-2}$, corresponding to a global significance of $1.8\,\sigma$.

\section{Conclusions}
\label{sec:conclusion}
A search has been performed for the standard model Higgs boson decaying
into two photons using data obtained from pp collisions at $\sqrt{s}=7\TeV$
corresponding to an integrated luminosity of 4.8\fbinv.
The selected events are subdivided into classes
according to indicators of mass resolution and signal-to-background ratio,
and the results of a search in each class are combined.
The expected exclusion limit at 95\% confidence level is between
1.4 and 2.4 times the standard model cross section in the mass range
between 110 and 150\GeV.
The analysis of the data excludes at 95\% confidence level the standard model
Higgs boson decaying into two photons in the mass range 128 to 132\GeV.
The largest excess of events above the expected standard model background is
observed for a Higgs boson mass hypothesis of 124\GeV with a local
significance of $3.1\,\sigma$.
The global significance of observing an excess with a local
significance ${\geq}3.1\,\sigma$ anywhere in
the search range 110--150\GeV is estimated to be $1.8\,\sigma$.
More data are required to ascertain the origin of this excess.
\newpage

\section*{Acknowledgements}
\label{sec:acknowledgements}
We wish to congratulate our colleagues in the CERN accelerator departments for the excellent performance of the LHC machine. We thank the technical and administrative staff at CERN and other CMS institutes, and acknowledge support from: FMSR (Austria); FNRS and FWO (Belgium); CNPq, CAPES, FAPERJ, and FAPESP (Brazil); MES (Bulgaria); CERN; CAS, MoST, and NSFC (China); COLCIENCIAS (Colombia); MSES (Croatia); RPF (Cyprus); Academy of Sciences and NICPB (Estonia); Academy of Finland, MEC, and HIP (Finland); CEA and CNRS/IN2P3 (France); BMBF, DFG, and HGF (Germany); GSRT (Greece); OTKA and NKTH (Hungary); DAE and DST (India); IPM (Iran); SFI (Ireland); INFN (Italy); NRF and WCU (Korea); LAS (Lithuania); CINVESTAV, CONACYT, SEP, and UASLP-FAI (Mexico); MSI (New Zealand); PAEC (Pakistan); SCSR (Poland); FCT (Portugal); JINR (Armenia, Belarus, Georgia, Ukraine, Uzbekistan); MON, RosAtom, RAS and RFBR (Russia); MSTD (Serbia); MICINN and CPAN (Spain); Swiss Funding Agencies (Switzerland); NSC (Taipei); TUBITAK and TAEK (Turkey); STFC (United Kingdom); DOE and NSF (USA). Individuals have received support from the Marie-Curie programme and the European Research Council (European Union); the Leventis Foundation; the A. P. Sloan Foundation; the Alexander von Humboldt Foundation; the Belgian Federal Science Policy Office; the Fonds pour la Formation \`a la Recherche dans l'Industrie et dans l'Agriculture (FRIA-Belgium); the Agentschap voor Innovatie door Wetenschap en Technologie (IWT-Belgium); and the Council of Science and Industrial Research, India.

\bibliography{auto_generated}   

\cleardoublepage \appendix\section{The CMS Collaboration \label{app:collab}}\begin{sloppypar}\hyphenpenalty=5000\widowpenalty=500\clubpenalty=5000\input{HIG-11-033-authorlist.tex}\end{sloppypar}
\end{document}

%% file: HIG-11-033-authorlist.tex
\textbf{Yerevan Physics Institute,  Yerevan,  Armenia}\\*[0pt]
S.~Chatrchyan, V.~Khachatryan, A.M.~Sirunyan, A.~Tumasyan
\vskip\cmsinstskip
\textbf{Institut f\"{u}r Hochenergiephysik der OeAW,  Wien,  Austria}\\*[0pt]
W.~Adam, T.~Bergauer, M.~Dragicevic, J.~Er\"{o}, C.~Fabjan, M.~Friedl, R.~Fr\"{u}hwirth, V.M.~Ghete, J.~Hammer\cmsAuthorMark{1}, M.~Hoch, N.~H\"{o}rmann, J.~Hrubec, M.~Jeitler, W.~Kiesenhofer, M.~Krammer, D.~Liko, I.~Mikulec, M.~Pernicka$^{\textrm{\dag}}$, B.~Rahbaran, C.~Rohringer, H.~Rohringer, R.~Sch\"{o}fbeck, J.~Strauss, A.~Taurok, F.~Teischinger, P.~Wagner, W.~Waltenberger, G.~Walzel, E.~Widl, C.-E.~Wulz
% ...
\vskip\cmsinstskip
\dag:~Deceased\\
1:~~Also at CERN, European Organization for Nuclear Research, Geneva, Switzerland\\
% ...
